\documentstyle[11lomcon,cite,graphicx]{article}

\begin{document}

\title{EVALUATION OF THE HIGHER TWIST CONTRIBUTION TO THE MOMENTS
OF PROTON STRUCTURE FUNCTIONS $F_2$ AND $g_1$}

\author{ \underbar{M.Osipenko}$^{1,2,}$ \footnote{e-mail: osipenko@ge.infn.it},
S.Simula$^{3,}$ \footnote{e-mail: simula@roma3.infn.it},
G.Ricco$^{1,4}$,
G.Fedotov$^{2}$,
E.Golovach$^{2}$,
B.Ishkhanov$^{2}$,
E.Isupov$^{2}$,
V.Mokeev$^{5,2}$}

\address{$^1$Istituto Nazionale di Fisica Nucleare, Sezione di Genova, 16146 Genova, Italy}
\address{$^2$Skobeltsyn Institute of Nuclear Physics, 119992 Moscow, Russia}
\address{$^3$Istituto Nazionale di Fisica Nucleare, Sezione Roma III, 00146 Roma, Italy}
\address{$^4$Dipartimento di Fisica dell'Universit\`a, 16146 Genova, Italy}
\address{$^5$Thomas Jefferson National Accelerator Facility, Newport News, Virginia 23606}


\maketitle\abstracts{ We performed the measurement of the inclusive electron
scattering off the proton~\cite{osipenko,f2data} in the resonance region ($W<2.5$~GeV) at
momentum transfer $Q^2$ below $4.5$~(GeV/c)$^2$ with the CLAS detector.
The large acceptance of CLAS provided an access to a large,
continuous two-dimensional kinematic domain in $Q^2$ and $x$,
allowing therefore an integration of the data at fixed $Q^2$ over
$x$-interval. The covered $x$-interval at each measured $Q^2$ value
is sufficient for an evaluation of the higher moments ($n>2$).
From these data we extracted the structure function $F_2$ and,
by including other world data, we studied the $Q^2$ evolution of its moments,
$M_n(Q^2)$, in order to estimate the higher twist contributions.
A similar experiment with polarized proton target is completed at CLAS~\cite{fatemi}.
These new data allow an accurate determination of higher moments of the proton
structure function $g_1$. A preliminary phenomenological analysis~\cite{Simula}
indicates an excess of the higher twist contribution in the spin-dependent
structure function with respect to the spin-independent one.}

\section{Introduction}\label{sec:1}
 Investigation of the nucleon internal structure with electromagnetic
probes provided most striking success of the strong interaction theory, QCD.
Measured nucleon structure
functions give an access to both parton momentum distributions as well as
to the scale dependence of the parton coupling with photon. Here the former
quantities, parton momentum distributions within the nucleon, are purely
phenomenological observable, not derived from the first principles of QCD.
Meanwhile, the scale dependence is completely determined by QCD evolution
equations. Therefore, in order to compare directly QCD predictions
on the nucleon structure to a measurement one has to study the scale
dependence of the structure functions avoiding the problem of describing
parton momentum distributions, as it was proposed in Ref.~\cite{conf,Ricco1}.
This can be performed by measuring
moments of the structure functions in experiment and studying their
scale evolution. In the case of electron-proton scattering it implies
a measurement of $Q^2$ evolution of the moments of proton structure functions
$F_2$, $g_1$, $F_1$ and $g_2$.

 The method of studying moments is based on Operator Product Expansion (OPE)
of the virtual photon-nucleon scattering amplitude. This leads to the
description of the complete $Q^2$ evolution of the moments of the
nucleon structure functions. For example, n-th Cornwall-Norton non-singlet moment
of the (asymptotic) structure function $F_2(x,Q^2)$ for a massless nucleon
can be expanded as:
\begin{equation}\label{eq:1}
M^{CN}_n(Q^2)=\sum_{\tau=2k}^{\infty}E_{n \tau}(\mu,Q^2)
O_{n \tau}(\mu)\biggl(\frac{\mu^2}{Q^2}\biggr)^{{1 \over 2}(\tau-2)},
\end{equation}
\noindent where $k=1,2,...,\infty$, $\mu$ is the factorization scale,
$O_{n \tau}(\mu)$ is the reduced matrix element of the local operators
with definite spin $n$ and twist $\tau$, related to
the non-perturbative nucleon structure.  $E_{n \tau}(\mu,Q^2)$
is a dimensionless coefficient function describing the small distance behavior,
which can be perturbatively expressed as a power series in the running
coupling constant $\alpha_s(Q^2)$.

 In order to investigate the double expansion in Eq.~\ref{eq:1} we
truncated both series: in the running coupling constant $\alpha_s(Q^2)$
up to Next to Leading Order (NLO)
and in twists $\tau$, suppressed by a power of
$(\mu^2/Q^2)^{{1 \over 2}(\tau-2)}$, up to $\tau=6$ term.
This choice is limiting the analysis to the
kinematic region where these parameters are small. We fixed
the value of local operator $O_{n \tau}(\mu)$ matrix element
at large $Q^2$, where all next-to-leading twist expansion terms
were neglected. After that we explored low-$Q^2$ region to determine
the contribution of the higher twist terms.

\begin{figure}
\begin{center}
\includegraphics[bb=1cm 13.5cm 20cm 19cm, scale=0.45]{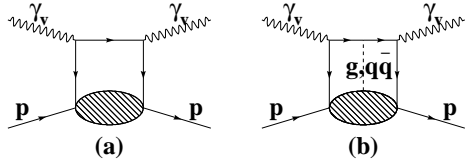}
\caption{\label{fig:TwistDiag} Twist diagrams:(a) the leading
twist contribution evaluated at leading order of pQCD; (b) the
contribution of higher twists, where current quark and nucleon remnant can
exchange by a system of particles consisting of gluons and $q\bar{q}$-pairs
whose complexity is increasing with twist order.}
\end{center}
\end{figure}

 The higher twists are related to quark-quark and quark-gluon
correlations, as illustrated by Fig.~\ref{fig:TwistDiag},
and should become important at small $Q^2$. In contrast to
the asymptotically free quarks, the quarks interacting among themselves
during the short time of the photon-proton scattering produce
the higher twist terms.
The importance of studying the multiparton correlations is due to
the fact that they are responsible for the phenomenon
of confinement and for the dynamical origin of proton mass.

 The recent experiments undertaken in Thomas Jefferson National Accelerator
Facility (USA) provide a large amount of data in the energy range not
well explored previously. Measurements of inclusive electron scattering
off polarized~\cite{fatemi} and unpolarized~\cite{osipenko,f2data} proton targets in Hall B
with maximum beam energy of 4.5 GeV allowed a precise extraction of the
moments of the proton structure functions $F_2$ and $g_1$ in the range of
momentum transfer, $Q^2$, from 0.1 up to 4.5~(GeV/c)$^2$. These
and complementary data from Deep Inelastic Scattering (DIS) were
analyzed in terms of perturbative QCD (pQCD) evolution equations and
OPE to create a complete picture of the proton structure at different scales.

\section{Experimental Moments of the structure functions}\label{sec:2}
 At small $Q^2$ values the moments contain non-negligible mass-dependent terms that
produce in Eq.~\ref{eq:1} additional $M^2/Q^2$ power corrections (kinematic twists),
that mix with dynamical twists under the renormalization-group equations.
To avoid these terms, the moments $M^{CN}_n(Q^2)$ have to be replaced
by the corresponding Nachtmann moments $M^{N}_n(Q^2)$
which by the construction allow to keep the form of the twist
expansion in Eq.~\ref{eq:1} (see also Ref.~\cite{rob}).
The Nachtmann moments of the structure function $F_2$ given by:
\begin{equation}\label{eq:2}
M^{N}_n(Q^2) = \int_0^1 dx \frac{\xi^{n+1}}{x^3} F_2(x,Q^2)
\Biggl[\frac{3+3(n+1)r+n(n+2)r^2}{(n+2)(n+3)}\Biggr],
\end{equation}
\noindent
where $r = \sqrt{1+4M^2x^2/Q^2}$ and $\xi = 2x/(1+r)$.
In the polarized case situation is more involved since
Nachtmann moments of the structure function $g_1$
depend on both $g_1$ and $g_2$:
\begin{eqnarray}\label{eq:3}
M^{N}_n(Q^2) = &&\int_0^1 dx \frac{\xi^{n+1}}{x^2} \Biggl\{g_1(x,Q^2)
\Biggl[\frac{x}{\xi}-
\frac{n^2}{(n+2)^2}\frac{M^2 x^2}{Q^2}\frac{\xi}{x}\Biggr]- \\ \nonumber
&& g_2(x,Q^2)\frac{M^2 x^2}{Q^2}\frac{4n}{n+2}\Biggl\},
\label{eq:i_nm1}
\end{eqnarray}
\noindent
Because of lack of experimental data on $g_2$, moments of
the polarized structure function $g_1$
contain an intrinsic unavoidable model dependence.

 The evaluation of experimental moment $M_n$ involves the
computation at fixed $Q^2$ of an integral over $x$ from the
structure function $F_2$ weighted with n-th power of $x$.
The integral over $x$ was computed numerically.
In $x$ intervals where data coverage was not complete
we applied essentially model independent interpolation method to avoid
as much as possible assumptions on $x$-shape of corresponding
momentum distributions. This was accomplished by normalizing the interpolation
function directly to experimental data located at the edges of
interpolating interval, independently for each $Q^2$ value.
Therefore, the obtained $Q^2$ evolution of the moments is free
of any model assumptions on the interpolating parton
momentum distributions. The low-$x$ extrapolation has been handled
by using two parameterizations and estimating the difference as
the systematic error. The error come out very small, thanks to the low-$x$
data from HERA. One can note that the low-$x$ extrapolation is
only important for the lowest moment, while in higher moments,
which are of the main interest of this analysis, the low-$x$ part
is strongly suppressed by a power of $x$. The analysis of the proton
structure function $g_1$ was based not on experimental data, but
on a parameterization. A more careful data-based analysis is
underway now and results obtained here are therefore preliminary.

 The moments were obtained with remarkable statistical and systematic
precision of the order of a few percent. In particular, higher moments
($n>2$) have almost 100\% of significant $x$-interval covered by
high precision CLAS data and therefore have tiny error bars. This
also allowed to extract the value of QCD running coupling constant
$\alpha_S(M^2_Z)$ with good precision~\cite{alphas}.

\section{OPE analysis}\label{sec:3}
 The experimental Nachtmann moments were analyzed in terms of the
following twist expansion
\begin{equation}\label{eq:twists}
M_n^N(Q^2) = LT_n(Q^2) + HT_n(Q^2) ~ ,
\end{equation}
\noindent where $LT_n(Q^2)$ is the leading twist moment and
$HT_n(Q^2)$ is the higher-twist contribution given by~\cite{Ji}
\begin{equation}\label{eq:HT}
HT_n(Q^2)=
a_n^{(4)}\biggl[\frac{\alpha_s(Q^2)}{\alpha_s(\mu^2)}\biggr]^{
\gamma_n^{(4)}}\frac{\mu^2}{Q^2}
+a_n^{(6)}\biggl[\frac{\alpha_s(Q^2)}{\alpha_s(\mu^2)}\biggr]^{
\gamma_n^{(6)}}\frac{\mu^4}{Q^4} ~ ,
\end{equation}
\noindent here the logarithmic pQCD evolution of the twist-$\tau$
contribution is accounted for by the term of LO-wise form
with an effective anomalous dimension $\gamma_n^{(\tau)}$
and the matrix element $a_n^{(\tau)}$ ($O_{n \tau}(\mu)$ in Eq.~\ref{eq:1})
fixes normalization of the twist-$\tau$ term at large $Q^2$.

The leading twist $LT_n(Q^2)$ term was calculated in pQCD to NLO
as the sum of a non-singlet and singlet terms.
Using the decoupling feature in the pQCD evolution~\cite{Ricco1}
of the singlet quark and gluon densities at large $x$ we
considered a pure non-singlet evolution for $n > 2$.
Therefore, for $n > 2$ leading twist contain one unknown parameter,
the matrix element $O_{n \tau}(\mu)$.
In order to (partly) account for the higher perturbative orders of pQCD
we used Soft Gluon Resummation (SGR) technique as in~\cite{SIM00}.
The resummation of soft gluons does not introduce any further
parameter in the description of the leading twist.
Leading twist normalization parameter as well as the higher-twist parameters
$a_n^{(4)}, \gamma_n^{(4)}, a_n^{(6)}, \gamma_n^{(6)}$, were
simultaneously determined in a $\chi^2$-minimization procedure
and reported in Refs.~\cite{osipenko,Simula}.

\begin{figure}
\begin{center}
\includegraphics[bb=1cm 7cm 20cm 19cm, scale=0.25]{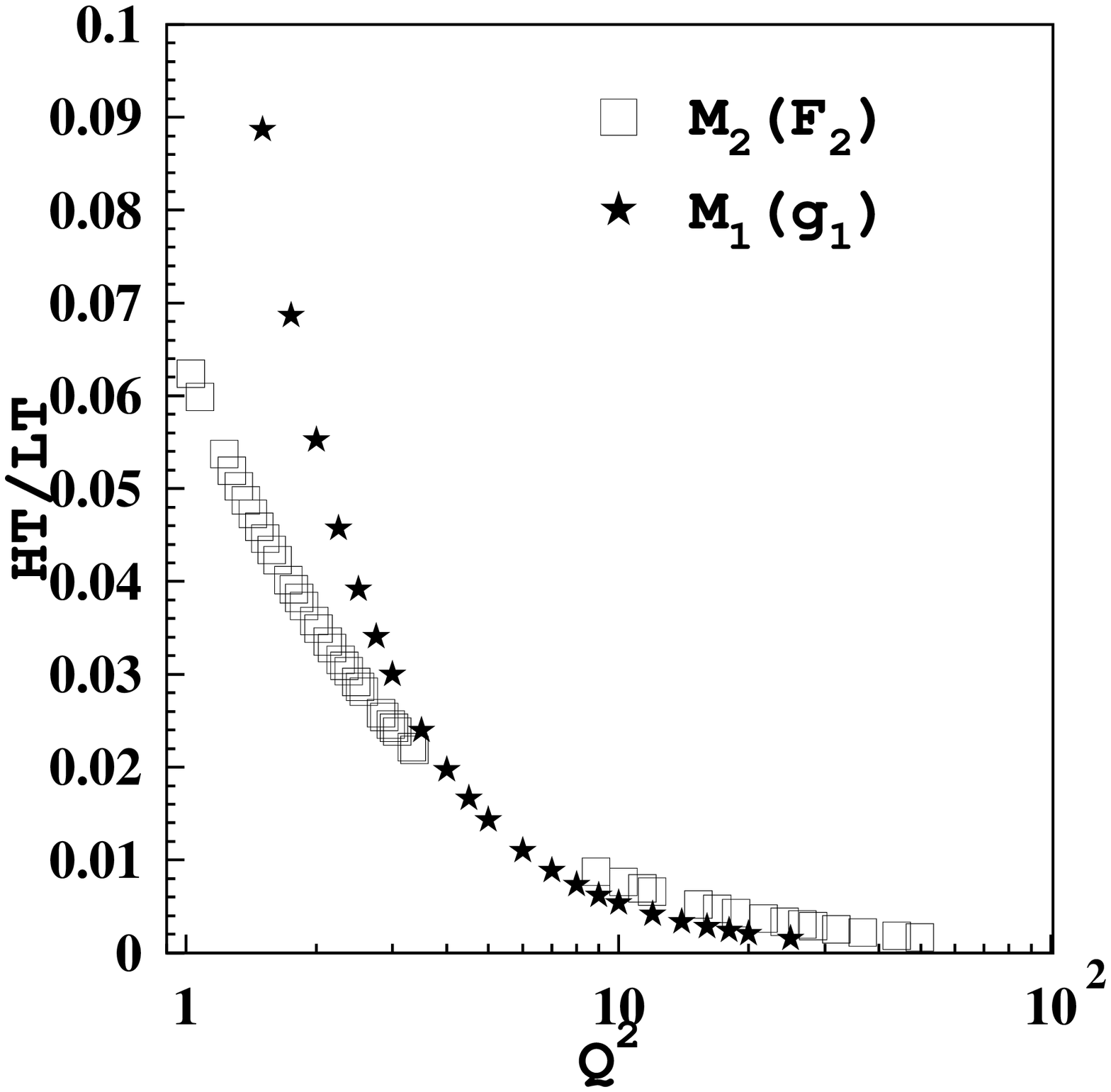}%
\includegraphics[bb=1cm 7cm 20cm 19cm, scale=0.25]{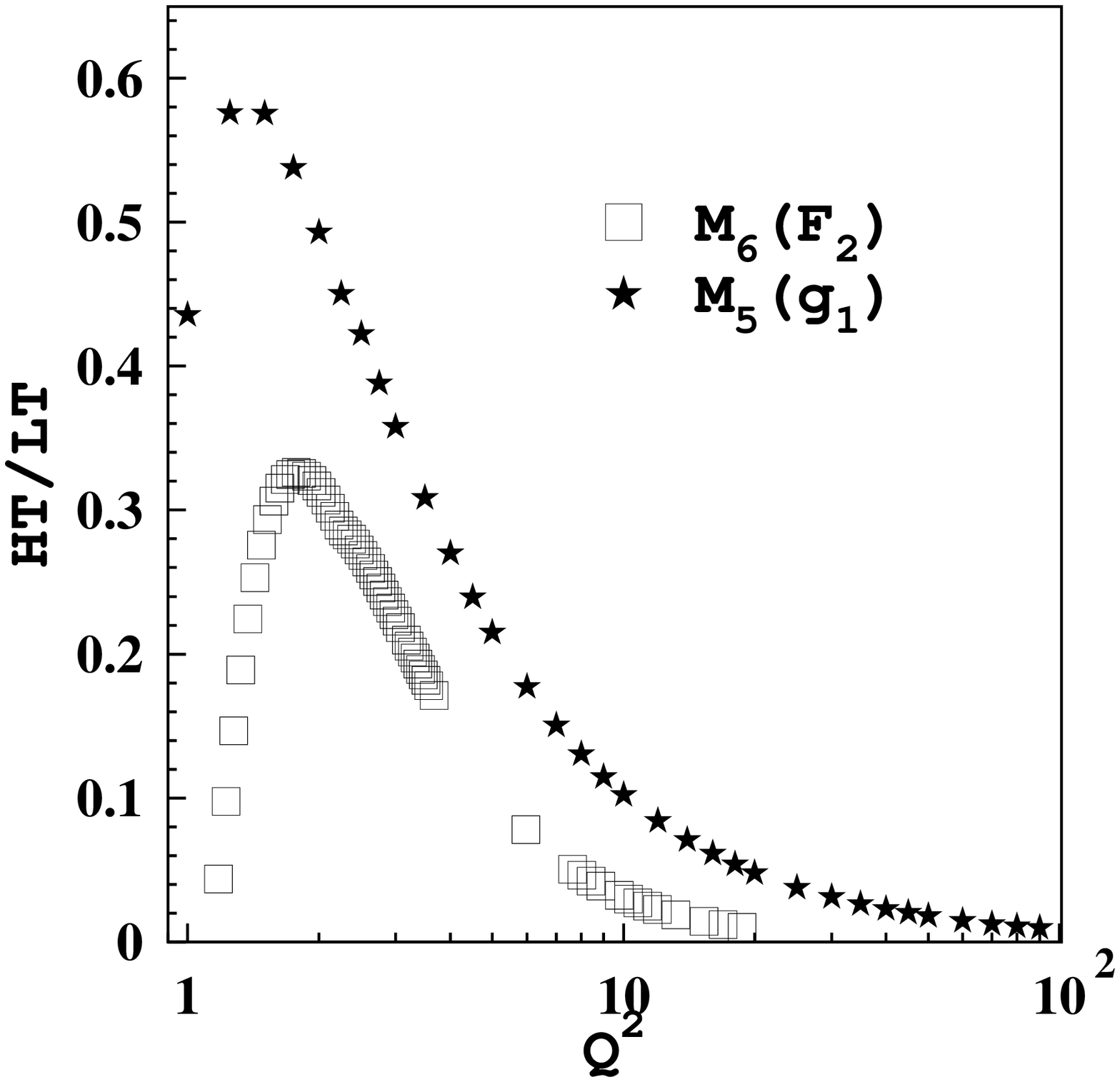}
\caption{\label{fig:2} Ratio of the higher to leading twist
contributions to the moments ($q(x)$ left panel and $x^4 q(x)$ right panel)
of the proton structure functions $F_2$ and $g_1$.}
\end{center}
\end{figure}

The obtained results can be summarized as follows:
~1) the contribution of the leading twist remains dominant down to
$Q^2$ of the order of a few (GeV/c)$^2$;
~2) different higher twist terms tend to compensate each other
in such a way that their sum is small even in a $Q^2$ region
where their absolute contributions exceed the leading twist (for details see Ref.~\cite{osipenko});
3) the contribution of higher twists relative to the leading one
is very sensitive to the parton polarization. This can be seen in the comparison
of the ratio higher to leading twists for structure functions $F_2$ and $g_1$
shown in Fig.~\ref{fig:2}. The power of $x$ in the moment is the same for both
structure functions, but the ratio is different by almost factor of two at
low $Q^2$. For the higher $Q^2$ values the lowest moment of both
structure functions becomes very similar. It gives an idea that
the enhancement of the higher twist contribution in $g_1$ moments
is due to presence of $P_{33}(1232)$ resonance. This excited nucleon state
gives strongly negative contribution to the total structure function $g_1$
breaking the quark-hadron duality expectations,
but it quickly disappears with rising $Q^2$ because of the rapid fall-off of the
$P_{33}(1232)$ form-factor.

\end{document}